\documentclass[prl,aps,showpacs,twocolumn]{revtex4-1}
\usepackage{graphicx}
\usepackage{amsmath,amssymb,color}
\usepackage[english]{babel}
\parskip=\medskipamount

\newcommand{\eq}[1]{(\ref{#1})}
\newcommand{\fig}[1]{Fig.\ref{#1}}

\newcommand{\be}{\begin{equation}}
\newcommand{\ee}{\end{equation}}

\begin{document}

\title{Effects of topological constraints on globular polymers}

\author{Maxim V. Imakaev,\textit{$^{a}$},
Konstantin M. Tchourine,\textit{$^{b}$},
Sergei K. Nechaev, $^{\ast}$\textit{$^{c,d,e,}$},
and  Leonid A. Mirny $^{\ast}$\textit{$^{a,f}$}}

\affiliation{$^a$Department of Physics, MIT, Cambridge, MA 02139, USA; leonid@mit.edu\\
$^b$Center for Genomics and Systems Biology, New York University, New York, USA  \\
$^c$Universit\'e Paris-Sud/CNRS, LPTMS, UMR8626, B\^at. 100, 91405 Orsay, France; sergei.nechaev@gmail.com \\
$^d$P.N.Lebedev Physical Institute of the Russian Academy of Sciences, 119991, Moscow, Russia \\
$^e$ National Research University Higher School of Economics, 109028 Moscow, Russia\\
$^f$Institute for Medical Engineering and  Science, MIT, Cambridge, MA 02139, USA}

\begin{abstract}
Topological constraints can affect both equilibrium and dynamic properties of polymer systems, and
can play a role in the organization of chromosomes. Despite many theoretical studies, the effects of
topological constraints on the equilibrium state of a single compact polymer have not been
systematically studied. Here we use simulations to address this longstanding problem. We find that
sufficiently long unknotted polymers differ from knotted ones in the spatial and topological states
of their subchains. The unknotted globule has subchains that are mostly unknotted and form
asymptotically compact $R_G(s) \sim s^{1/3}$ crumples. However, crumples display high fractal
dimension of the surface $d_b = 2.8$, forming excessive contacts and interpenetrating each other.
We conclude that this topologically constrained equilibrium state resembles a conjectured crumpled
globule  [Grosberg \textit{et al., Journal de Physique}, 1988, \textbf{49}, 2095], but differs from
its idealized hierarchy of self-similar, isolated and compact crumples. 
\end{abstract}

\maketitle

\section{Introduction}

Topological constraints, i.e. the inability of chains to pass through each other, have significant
effects on both equilibrium and dynamic properties of polymer systems
\cite{gel-electr,old-dna,gr-rab}  and can play important roles in the organization of chromosomes
\cite{halverson2014melt, gr-rab,stasiak,leonid_review}. Previous theoretical studies  suggested that
topological constraints {\it per se}  compress polymer rings or polymer subchains by topological
obstacles imposed by surrounding subchains \cite{gns,kh-nech,cates_deutsch}.  This compression
makes a subchain of length $s$ form a space-filling configuration that has an average radius of
gyration $R_G(s) \sim  s^{1/3}$. Recent simulations of topologically constrained unconcatenated
polymer rings in a melt \cite{reith2011gpu,gr-krem1,gr-krem2,suzuki2009dimension,rosa2013ring} have
demonstrated the effect of compression into space-filling configurations and confirmed $s^{1/3}$
scaling, thus providing strong support to previous conjectures.

\begin{figure}
\includegraphics[width=8cm]{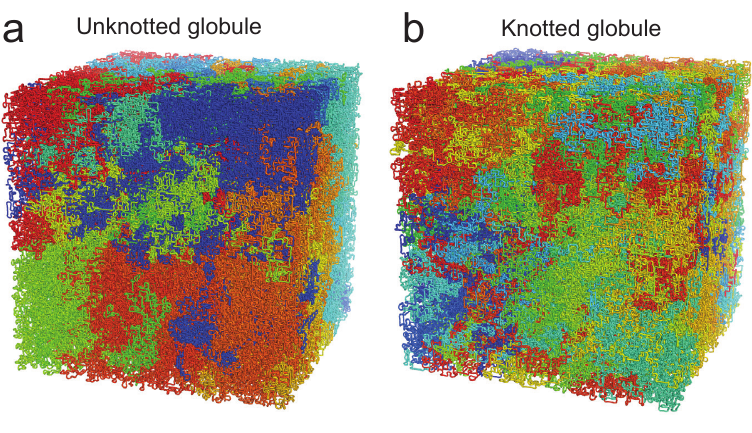}
\caption{{\bf Representative conformations of two different types of globules} (confined polymer rings, length $N=256\,000$): (a)
the unknotted globule formed by a single polymer ring with topological constraints; (b) the knotted globule formed by a polymer ring without topological constraints. Both chains are painted in red-yellow-green-blue along the polymer length.}
\label{fig:01}
\end{figure}

The role of topological constraints in the {\em equilibrium} state of a single compact and
unknotted  polymer remains unknown. Previous studies \cite{gns,kh-nech,gr-rab} have put forward a
concept of the {\em crumpled globule} as the equilibrium state of a compact and unknotted polymer.
In the crumpled globule, the subchains were suggested to be space-filling and unknotted. This
conjecture remained untested for the quarter of the century. Here, we  test this conjecture by
comparing equilibrium compact states of a topologically constrained and unknotted polymer, referred
to below as the {\it unknotted globule}, with those a topologically relaxed one, referred to below
as the {\it knotted globule} (Fig. \ref{fig:01}).

Recent computational studies examined the role of topological constraints in the {\em
non-equilibrium} (or quasi-equilibrium) polymer states that emerge upon polymer collapse
\cite{lieberman, rost,obukhov2,chu,chertovich2014crumpled}. This non-equilibrium state, often
referred to as the {\em fractal globule} \cite{lieberman,leonid_review}, can indeed possess some
properties of the conjectured equilibrium crumpled globule. The properties of the fractal globule, its
stability \cite{schiessel}, and its connection to the equilibrium state are yet to be understood.

Elucidating the role of topological constraints in equilibrium and non-equilibrium polymer systems
is important for understanding the organization of chromosomes. Long before experimential data on
chromosome organization became available \cite{lieberman}, the crumpled globule was suggested as a
state of long DNA molecules inside a cell \cite{gr-rab}.  Recent progress in microscopy
\cite{cremer2011review} and genomics \cite{dekker_NRG} provided new data on chromosome organization
that appear to share several  features with topologically constrained polymer systems
\cite{rosa_everaers,lieberman,grosberg2012review}. For example, segregation of chromosomes into
territories resembles segregation of space-filling rings \cite{gr-krem2,stasiak}, while features of
intra-chromosomal organization revealed by Hi-C technique are consistent with a non-equilibrium
fractal globule emerging upon polymer collapse \cite{lieberman,leonid_review,machine} or upon
polymer decondensation \cite{rosa_everaers}. These findings suggest that topological constraints
can play important roles in the formation of chromosomal architecture \cite{halverson2014melt}.

Here we examine the role of topological constraints in the equilibrium state of a compact polymer
(\fig{fig:01}). We perform equilibrium Monte Carlo simulations of a confined unentangled polymer
ring with and without topological constraints. Without topological constraints, a polymer forms a
classical equilibrium globule with a high degree of knotting \cite{lgk,knots-kardar}. A polymer is
kept in the globular state by impermeable boundaries, rather than pairwise energy interactions,
allowing fast equilibration at a high volume density.

We find that topological states of closed subchains (loops) are drastically different in the two
types of globules and reflect the topological state of the whole polymer. Namely, loops of the
unknotted globule are only weakly knotted and mostly unconcatenated. We also find that spatial
characteristics of small knotted and unknotted globules are very similar, with differences starting
to appear only for  sufficiently large globules. Subchains of these large unknotted globules become
asymptotically compact ($R_G(s)\sim s^{1/3}$), forming crumples. Analyses of the fractal dimension
of surfaces of loops suggest that crumples form excessive contacts and interpenetrate each other.
Overall, the asymptotic behavior we find support the conjectured crumpled globule concept
\cite{gns}. However, our results also demonstrate that  the internal organization of the unknotted
globule at equilibrium differs from an idealized hierarchy of self-similar isolated compact
crumples.

\section{Results}

\subsection{Model}

A single homopolymer ring  with excluded volume interactions was modeled on a cubic lattice and confined into a cubic container at a volume density $0.5$. The Monte Carlo method with non-local moves \cite{mc-moves} allowed us to study chains up to $N=256\,000$.
If monomers were prohibited to occupy the same site, this Monte Carlo move set naturally constrains topology, and the polymer remains unknotted. Setting a small finite probability for two monomers to occupy the same lattice site would let two regions of the chain cross. This would relax topological constraints while largely preserving the excluded volume (\fig{fig:Co-occupation}). The topological state of a loop was characterized by $\varkappa$, the logarithm of the Alexander polynomial evaluated at $-1.1$ \cite{grosberg-knots,knots-kardar,virnau,kolesov2007protein}. To  ensure equilibration,  we estimated the scaling of the equilibration time with $N$ for $N\le 32\,000$, extrapolated it to large $N$, and ran simulations of longer chains, $N=108\,000$ and $256\,000$, to exceed the estimated equilibration time (see Supplement and Fig. \ref{fig:07} for details). We also made sure that chains with topological constraints remain completely unknotted through the simulations, while polymers with relaxed topological constraints become highly entangled \cite{knots-kardar} (Fig. \ref{fig:AlexKnotted})

\subsection{Topological properties}

\begin{figure}  
\includegraphics[width=8.2cm]{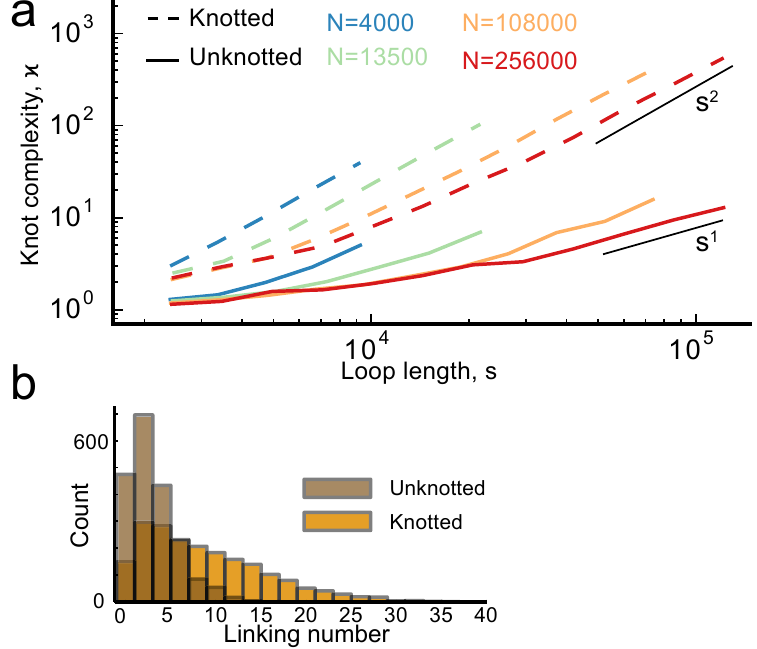}
\caption{{\bf Topological properties of polymer loops in the knotted and unknotted globules.}
(a) Knot complexity of polymer loops as a function of their length, $s$, for chains of different length $N$ (shown by colors) in  knotted (dashed) and unknotted (solid) globules. (b) Distribution of the
linking numbers for non-overlapping loops of length $s=9000$ to $11\,000$ in $32\,000$-long globules.}
\label{fig:02}
\end{figure}

First, we asked how the topological state of the whole polymer influences the topological properties of its
subchains. Because a topological state can be rigorously defined only for a closed contour, we
focused our analysis on loops, i.e. subchains with two ends occupying neighboring lattice sites.
Fig. \ref{fig:02}a presents the average knot complexity $\langle \varkappa(s) \rangle$ for loops
of length $s$  for both types of globules. We found that loops of the knotted globule
were highly knotted, with the knot complexity rising sharply with $s$. Loops of the
unknotted globule, on the contrary, were weakly knotted, and their complexity increased slowly with
length. Their knotting complexity was also very variable, indicating the abundance of slip knots \cite{sulkowska2012slipknots} (Fig. \ref{fig:slipknots32}).

This striking difference in the topological states of loops for globally
knotted and unknotted chains is a manifestation of the general statistical
behavior of so-called matrix--valued Brownian Bridges (BB) \cite{nechaev1996random}. The knot complexity
$\varkappa$ of  loops in the topologically unconstrained globule is expected to grow as $\varkappa(s)\sim s^2$.
In contrast, due to the global topological constraint imposed  on the unknotted globule, the knot complexity of its loops grows slower,  $\varkappa(s) \sim s$, which follows from the statistical behavior of BB in spaces of constant negative curvature (see Appendix,  \fig{fig:02}a, and \cite{nech_vas,nechaev1996random,nechaev1991limiting,furstenberg1963noncommuting} for details). Our simulations are in good agreement with this theory (\fig{fig:02}a).

Another topological property of loops of a globule is the degree of concatenation between the loops.
We computed the linking number for pairs of non-overlapping loops in each globule
(\fig{fig:02}b) and found that loops in the unknotted globule are much less concatenated than loops
in the knotted globule.

Taken together, these results show that the topological state of the whole (``parent'') chain propagates to the ``daughter'' loops. While loops of the unknotted globule are linked and
knotted, their degree of entanglement is much lower than for the loops in the topologically relaxed
knotted  globule (Fig. \ref{fig:slipknots32}).
Our results also demonstrate that loops of a single unknotted globule are not equivalent to recently studied rings in a melt, which were unknotted and unconcatenated. We further examine this parallel below.


\subsection{Spatial properties}

\begin{figure}[ht]
\includegraphics[width=8.5cm]{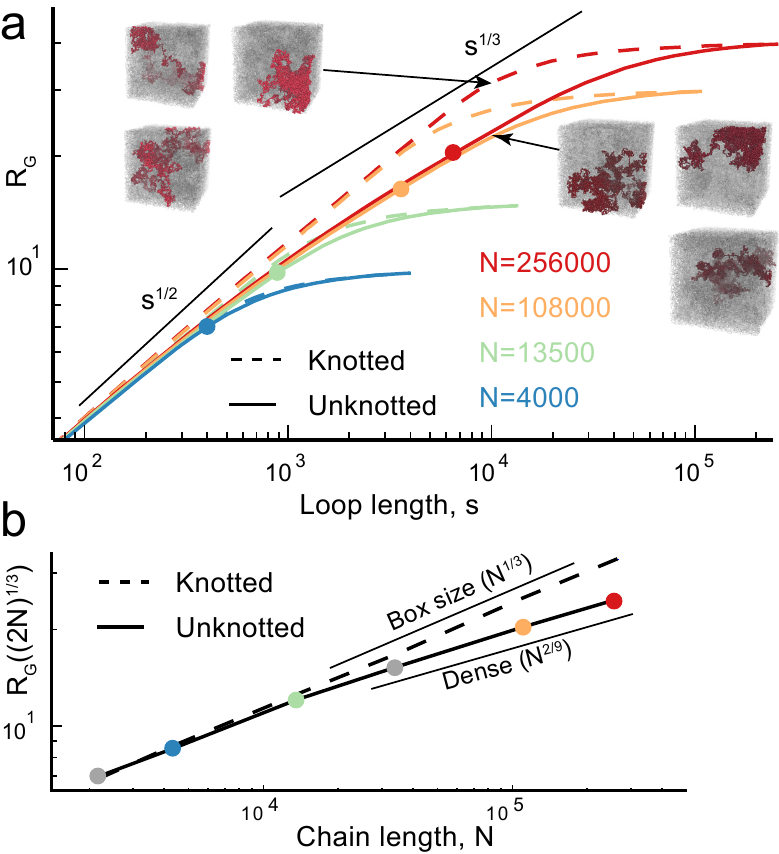}
\caption{
{\bf Spatial properties of loops in the knotted and unknotted globules.} (a) The average radius of
gyration ${R_G}(s)$ for loops of length $s$. Inset shows conformations of three $10\,000$-monomer loops
for $N=256\,000$ globules. Circles indicate subchains of length $(2\cdot N)^{2/3}$, which equals the squared size of the confining container.  (b) Dependence of $R_G(s=(2N)^{2/3})$ on the chain size $N$. Colored circles match to circles in (a); grey circles denote $N=2000$ and $N=32000$, which are not shown in (a)}
\label{fig:03}
\end{figure}

Next, we examined the effects of topological constraints on the spatial properties of loops. We computed an average gyration radius ${R_G}(s)$ (\fig{fig:03}) as a function of loop length $s$, in chains of different length $N$.  The two types of globules show different trends in $R_G(s)$.

The behavior of ${R_G}(s)$ for  knotted globules has two regimes that are well-known and described by the Flory theorem \cite{deGennes_book}. Shorter subchains behave as Gaussian coils $R_G(s) \sim s^{1/2}$ until they reach the confining walls at $R \sim N^{1/3}$, i.e. for $s \lesssim s_c \sim  N^{2/3}$. For longer subchains, $s \gtrsim s_c$, ${R_G}(s)$
plateaus at ${R_G}(s) \sim N^{1/3}$. Note that this is qualitatively similar to $R_G(s)$ for a phantom chain confined to a box (Fig. \ref{fig:Phantom}).  Overall, for knotted globules, our results are in line with theoretical predictions.

Subchains in unknotted globules were proposed to be compressed by topological constraints and follow ${R_G}(s) \sim s^{1/3}$ relationship \cite{gns}. Surprisingly, we do not observe any significant compression of loops in chains of length $N \leq13\,500$, as there is little difference between ${R_G}(s)$ curves for the two types of globule. For longer chains, $N \geq10^5$, we observe an increasing difference between $R_G(s)$ of knotted and unknotted globules. Little difference is seen in other moments of the distribution of subchain sizes (Fig. \ref{fig:kurtosis}). In the unknotted globules, we observe a range of subchain length, $10^3 \leq s \leq 10^4$, in which the subchains are compressed (${R_G}(s) \sim s^{1/3}$) and the curves for the two largest systems collapse. However, this scaling regime arises as a gradual decrease from $R_G(s) \sim s^{1/2}$ (see \fig{fig:fullpage}), and thus cannot be established unambiguously.

Previous studies have established that topological constraints become relevant for chains that are several times longer than a characteristic length $N_e$ called the enganglement length \cite{halverson2014melt,deGennes_book,GKh-book,everaers2004rheology}. For a similar system, an equilibrium melt of rings at the same volume density, it was estimated that $N_e \approx 175$ \cite{gr-krem1}, and topological constraints become relevant only  above several $N_e$, i.e. for $N \gtrapprox 1000$ \cite{gr-krem1}. Following this logic, we expect that in our system topological constraints become relevant for subchains of length $s \gtrapprox 1000$. However, subchains $s \geq s_c$ experience confinement, which overshaddows topological compression. Indeed, when we consider only loops not touching the boundary, we see the difference for subchains $s \gtrapprox 1000$ (Fig. \ref{fig:not_touching}). For topological constraints to be relevant, a polymer should have subchains that do not experience external confinement ($s \leq s_c\sim N^{2/3}$) but are sufficiently long to experience topological compression ($s \gtrapprox 1000$). This sets a lower limit for a polymer to experience topological constraints ($N^{2/3} \gtrsim 1000$). Consistently, we observe a difference between ${R_G}(s)$ for the two types of  globules for $N \gtrsim 32\,000$ (\fig{fig:03}a, \ref{fig:fullpage}).

\begin{figure}[ht]
\includegraphics[width=8.6cm]{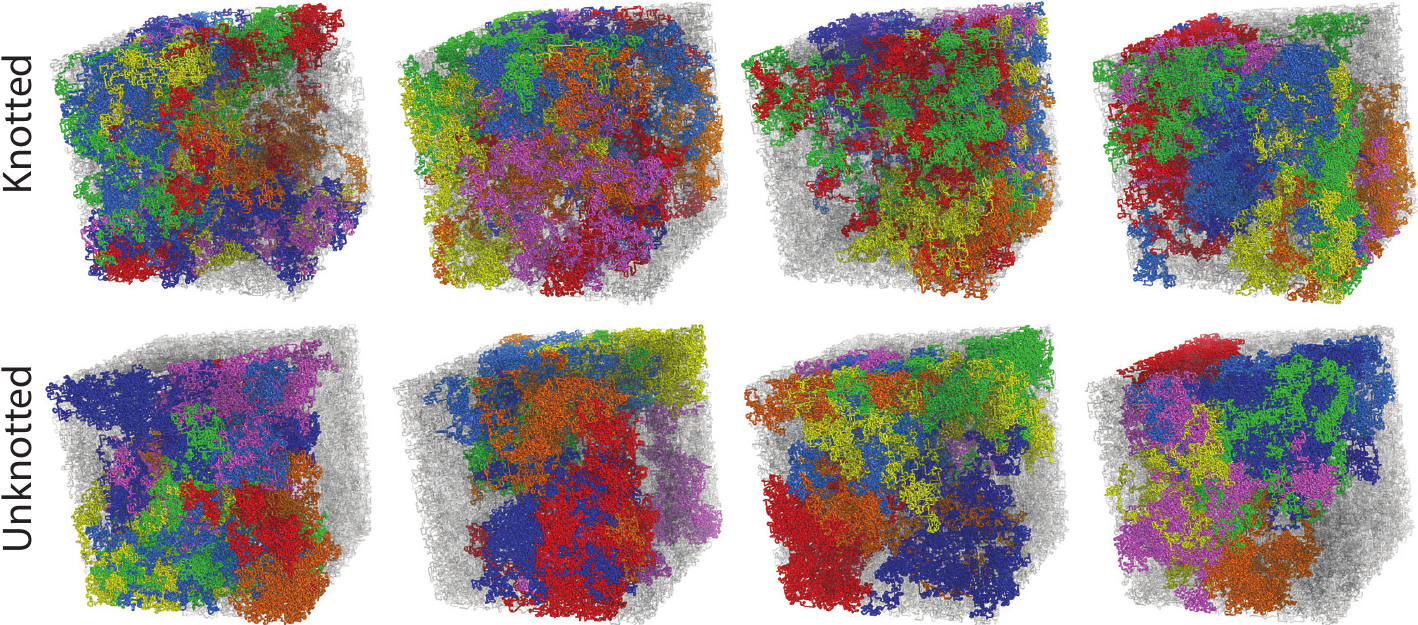}
\caption{
{\bf Compression of subchains in unknotted loops.} Seven consecutive subchains of length $s=10\,000$ are highlighted with consecutive colors (red, orange, yellow, green, marine, blue, and violet) in four knotted and  unknotted $N=256\,000$ globules; rest of the chain is shown in grey.  }
\label{fig:subchains}
\end{figure}

To test the conjecture that $R_G(s) \sim s^{1/3}$, we need to separate compression by topological constraints from the effect of confinement. To this end, we focused on loops of size $s = s_c$, which are the largest loops not affected by confinement.
Fig. \ref{fig:03}b presents ${R_G}(s_c)$ as a function of $N$ and clearly shows distinct scalings for knotted and unknotted globules.
The knotted globule follows $R_G(s_c) \sim N^{1/3}$, which is a consequence of $R_G(s) \sim s^{1/2}$. For the unknotted globule, however, we observe $R_G(s_c) \sim N^{2/9}$, which corresponds to $R_G(s) \sim s^{1/3}$.
Thus, in agreement with previous conjectures
\cite{gns,nech_vas}, topological constraints lead to the formation of "space-filling" subchains, i.e.
$R(s) \sim s^{1/3}$. However, the compressing effect of topological constraints becomes evident only
for very long polymers, such as $N \gtrapprox 10^5$. Space-filling crumples are visible in unknotted globules of length $N=256\,000$,  but the difference is visually subtle (\fig{fig:subchains}). Distinguishing individual knotted and unknotted globules by eye is challenging, but a pattern is visible when several globules are compared.

Our analysis reveals significant effects of global topological constraints on the topological and spatial
characteristics of loops. Next, we asked whether topological states and sizes of loops are intrinsically connected. We computed $R_G$ and
$\varkappa$ for loops of length $s=20\,000$ in $N=256\,000$ globules (Fig. \ref{fig:3c_suppl}).
Despite having similar ${R_G}$, loops from the two types of globules show
different knot complexities: all loops from the unknotted globule were significantly
less knotted than loops of the same length in the knotted globule.
Moreover, $\varkappa$ for loops in the knotted globule negatively correlates with $R_G$: more compact loops form more complex knots in the system were no topological constraints were present.
%
%
This relationship, however, does not hold across globules: loops in the unknotted globule are on average more compact and less knotted.
These observations highlight that there is no simple relationship between spatial and topological properties of closed contounrs.


\subsection{Contact probability}


Another important characteristic of internal polymer organization is the probability $P_c(s)$  of a contact
between two monomers separated by a contour length $s$.  For example, for a 3D random walk, $P_{c,RW}(s) \sim s^{-3/2}$ (see Supplemental Information).
A recently developed experimental technique, Hi-C, measures $P_c(s)$ experimentally for chromosomes inside cells
\cite{dekker_NRG,lieberman}. Comparison of experimental and theoretical $P_c(s)$ can shed light on polymer organization of chromosomes \cite{lieberman,naumova,le-caulobacter}. In our previous work, we found that a non-equilibrium fractal globule, which
emerges upon a polymer collapse, has $P_c(s) \sim s^{\alpha}$, $\alpha \approx -1$. The $P_c(s)$ scaling for the fractal globule agrees with $P_c(s)$ from the Hi-C data for human chromosomes better than other polymer ensembles \cite{lieberman}.

\begin{figure}[ht]
\includegraphics[width=8cm]{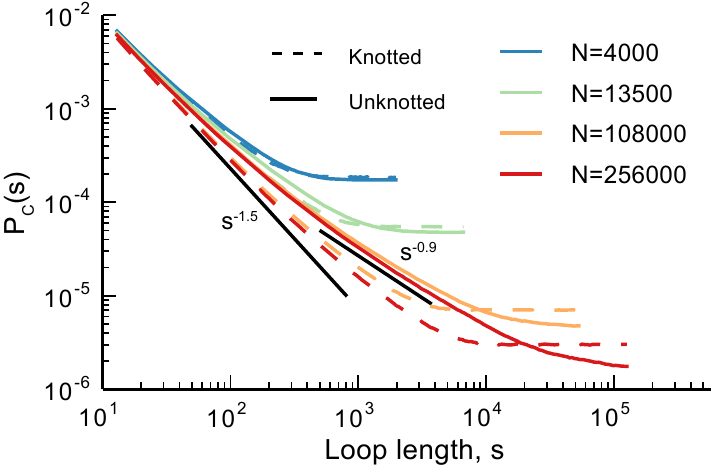}
\caption{ {\bf Scaling of the contact probability.} $P_c(s)$ is shown for knotted and unknotted globules of
different sizes. Compare to \fig{fig:03}a.}
\label{fig:04}
\end{figure}

Figure  \ref{fig:04}  presents $P_c(s)$ for the knotted and unknotted globules. For the
knotted globule, as expected, we observed two regimes $P_c(s) \sim s^{-3/2}$ for $s \lesssim s_c$,
followed by a plateau for $s \gtrsim s_c$ \cite{gr_bor}. As  above, an equilibrium globule without topological constraints can be
considered as a "gas of random walks" \cite{gr_bor}, i.e. short chains ($s \lesssim s_c$) behave as random walks. Different random walks are mixed and are equally likely to contact each other, leading to the plateau of $P_c(s)$ for $s \gtrsim s_c$.
Subchains in the unknotted globule, however, experience additional confinement by topological
constraints and have a different $P_c(s)$. For $N \leq 13\,500$,
little difference is observed between the two types of globules, which is consistent with our
observation that topological effects play little role for shorter polymers.

Longer unknotted globules show a different $P_c(s)$ curve with a
less steep decline of $P_c(s)$ for small $s$ and no distinct plateau for large $s$.
$P_c(s)$ plots and their derivatives (Fig. \ref{fig:fullpage}) suggest a possible
scaling regime $P_c(s) \sim s^\alpha$,   $-1  < \alpha < -0.8$ for loops of $s=10^3-10^4$, where topological constraints are expected to play a bigger role. The value  
observed for the melt of rings \cite{gr-krem1,gr-krem2}, $\alpha \approx -1.17$, is outside of this range highlighting a difference between these systems.

Note, however, that estimating scaling of $P_c(s)$ for both types of globules is challenging due to a broad transition between different regimes and the effects of confinement. As seen on Fig. \ref{fig:fullpage}, even for the knotted globule, where the scaling of $P_c(s) \sim s^{-3/2}$ is known, it can be observed only asymptotically.

\subsection{Fractal structure of loops}

Loops of the unknotted globule become asymptotically compact as the polymer size increases, forming crumples. The
question that follows is whether such crumples become more isolated from each other. To answer this question, we studied shapes of crumples formed by loops; we calculated the fractal dimension of their surface and corrected for finite-size effects.
For a loop, the surface area of the boundary, $A$, is defined as the number of monomers forming contacts with the rest of the polymer \cite{halverson2014melt}. The fractal dimension of the loop boundary, $d_b$, is defined by $A(s) \sim s^{d_b/3}$. Note that $d_b$ denotes fractal dimension of the boundary only in the compact subchain regime, $R_G \sim s^{1/3}$; for non-compact subchains it measures the scaling of the subchain boundary with subchain length.

\begin{figure}[ht]
\includegraphics[width=8cm]{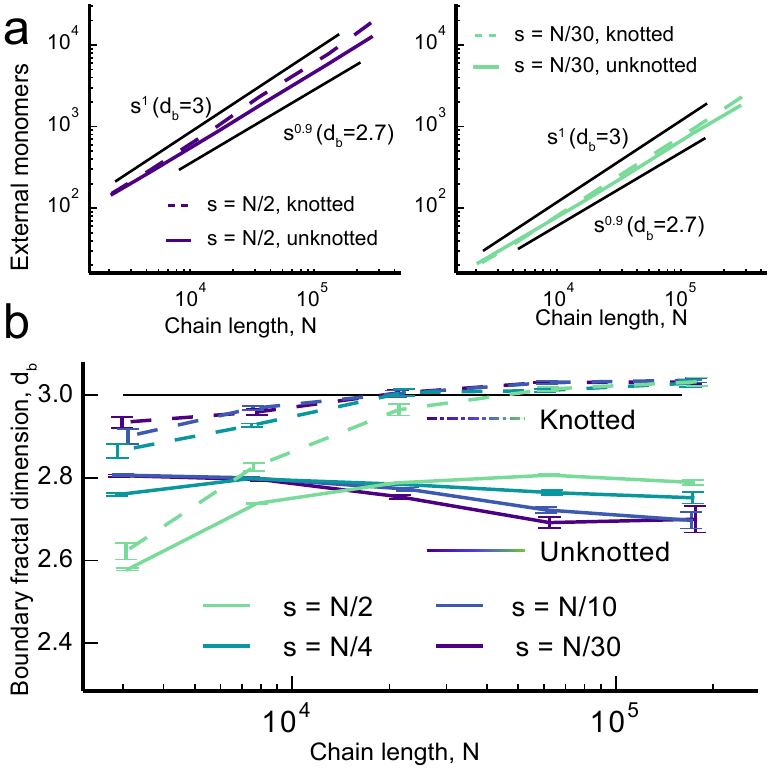}
\caption{{\bf Fractal dimension of a loop surface.}
\textbf{(a)} The surface area $A(s,N)$ of loops of length $N/2$ (left) and
$N/30$ (right) as a function of chain length $N$. \textbf{(b)} The fractal dimension of the surface measured from slopes in (a), as a function of $N$.
The slope in (a) is estimated from pairwise differences between
two neighboring values of $N$. Error bars show a standard error of the mean obtained by
bootstrapping. To allow comparisons between error bars, plots were shifted by a negligible offset
along the horizontal axis. }
\label{fig:05}
\end{figure}

Finite-size effects, i.e. effects of the global confinement on the loops, can be taken into account by a
function that depends on the fraction of the loop $s$ in the whole chain $N$,
$f\left(\frac{s}{N}\right)$, giving the surface area $A(s,N) =
f\left(\frac{s}{N}\right)s^{d_b/3} = g\left(\frac{s}{N}\right)N^{d_b/3}$.
We can then compute $d_b$ by comparing $A$ for chains with the same value of $s/N$, but in globules of different length $N$ (\fig{fig:05}a). Figure \ref{fig:05}b shows $d_b$ as a function of $N$ and
gives asymptotic behavior of $d_b$ for large $N$, where topological constraints become
most relevant. As expected, loops in the knotted globule have $d_b \approx 3$, suggesting that loops fully mix with
each other throughout their entire volume.
In the unknotted globule, however, loops have $d_b \approx 2.8$, indicating that loops are not fully mixed, yet not fully isolated.
A fractal dimension $d_b
=2$ would indicate interactions over two-dimensional surface area,
i.e. as bricks stacked together. The value of $d_b \approx 2.8$ comes close to the fractal dimension of a ring surface $\approx 2.85$ found for unconcatenated rings in a melt \cite{gr-krem2}. This indicates that, similar to unconcatenated rings, loops of the unknotted globule are not isolated, and form interdigitated compact crumples (\fig{fig:subchains}).

\section{Conclusions}

Our results provide strong support to the previous conjectures \cite{gns} and  reveal several new insights into the effects of topological constraints on the compact state of a polymer.
In agreement with \cite{gns}, we observe that topological constraints make a compact polymer acquire a new conformational state, previously called the crumpled globule. In this state, the unknotted polymer forms largely unknotted and weakly concatenated crumples. These crumples are asymptotically compact ($R_G(s) \sim s^{1/3}$) and differ from random walk subchains ($R_G(s) \sim s^{1/2}$) emerging in the absence of topological constraints. However, the effect is hard to detect, since large subchains experience confinement, while small subchains (less that a few $N_e$) do not feel topological constraints. As a result, only chains of some intermediate size ($10^3 \lesssim s \lesssim N^{2/3}$) form topologically compressed crumples. Similarly, the effects of topological constraints are most evident in large globules ($N \gtrsim 10^5$).

Unexpectedly, we also found that the loops in a globally unknotted polymer are somewhat knotted and concatenated. The brownian bridge argument explains this phenomenon and is in good quantitative agreement with  the scaling of $\varkappa(s)$ (Fig. \ref{fig:02}). Knots formed by loops of the unknotted globule are much less complex than those in the topologically unconstrained globule. Overall, this demonstrates how global topological constraints imposed on the whole chain propagate into local topological constraints acting on its subchains.

We found that the unknotted globule is not self-similar as a whole, but subchains show evidence of self-similarity for sufficiently large globules. Even the largest system considered ($N=256\,000$) shows a rather narrow (factor of $10$ in $s$) scaling regime in $R_G(s)$. Moreover, the fractal dimension of loop surfaces, $d_b \approx 2.8$, shows that compact crumples are neither fully isolated ($d_b=2$), nor fully mixed ($d_b=3$).  Some degree of mixing with neighboring subchains makes crumples swell, possibly narrowing the range of $s$ where subchains are self-similar, and further highlighting differences between a finite-size unknotted globule and an idealized hierarchical crumpled globule proposed theoretically \cite{gns}.

It is possible that features of the conjectured crumpled globule can be more evident in  a non-equilibrium state that emerges immediately after polymer collapse (often referred to as the fractal globule) \cite{lieberman,leonid_review}, rather than in the equilibrium system considered here.
In the non-equilibrium state, a broader regime of scaling in $P_c(s)$ and $R_G(s)$ suggests that even much shorter chains have self-similar organization (see also \cite{halverson2014melt}). We cannot rule out the possibility that effects other than than topological constraints, acting on shorter time scales, can constrain a collapsed chain in a transient state. This state would be different from the crumpled globule studied here.

We find many similarities and some notable differences between the unknotted globule formed by a single ring and the melt of unconcatenated rings  \cite{gr-krem1,gr-krem2}. Both systems show quantitatively similar compression of rings and loops by topological interactions, as follows from similar $R_G(s)$  asymptotic scalings and similar fractal dimension of the surface.
The systems however are different on many levels. While rings in a melt are monodispersed, unknotted and unconcatenated, loops of a single polymer have a broad size distribution, are knotted and concatenated. This variation in size can lead to swelling of larger loops. Moreover, larger loops experience global confinement of the globule. Topologically, the systems are different since the considered polymer ring has only one global topological constraint, while the melt has as many constraints as the number of rings. Nevertheless some characteristics of mid-size loops in unknotted globules resemble those of rings in a melt.

Overall, we find that equilibrium state of a single unknotted polymer chain is different from topologically relaxed system. It would be interesting to see whether and to what extent this phenomenon is observed in other physical systems where topology can play a role.


\section{Acknowledgments}
This work was supported by MIT-France Seed Funds and NCI-funded Center for Physical Sciences in
Oncology at MIT (U54CA143874). We are grateful to Geoffrey Fudenberg, Anton Goloborodko and
Christopher McFarland for many productive discussions and to Alexander Y. Grosberg for his feedback
and suggestions. S.N. acknowledges the support of the Higher School of Economics program for Basic
Research.

\bibliography{globules}{}


\section*{Appendix}

\section*{Statistics of matrix-valued Brownian Bridges }

The conditional distributions of a knot complexity of a subpart of a globally unknotted polymer
chain are a typical problem in the theory of Markov chains and deals with the determination of the
conditional probability for so-called \emph{Brownian Bridges} (BB). The investigation of statistics
of BB supposes the determination of the probability $P({\bf x},t|{\bf 0},T)$ for a random walk to
start at the point ${\bf x}={\bf 0}$, to visit the point ${\bf x}$ at some intermediate moment
$0<t<T$, and to return to the initial point ${\bf x}={\bf 0}$ at  the moment $T$. The same question
can be addressed for BB on the graphs of noncommutative groups, on Riemann surfaces and for
products of random matrices of groups \cite{nechaev1991limiting,nechaev1996random}.

\begin{figure}[ht]
\includegraphics[width=8cm]{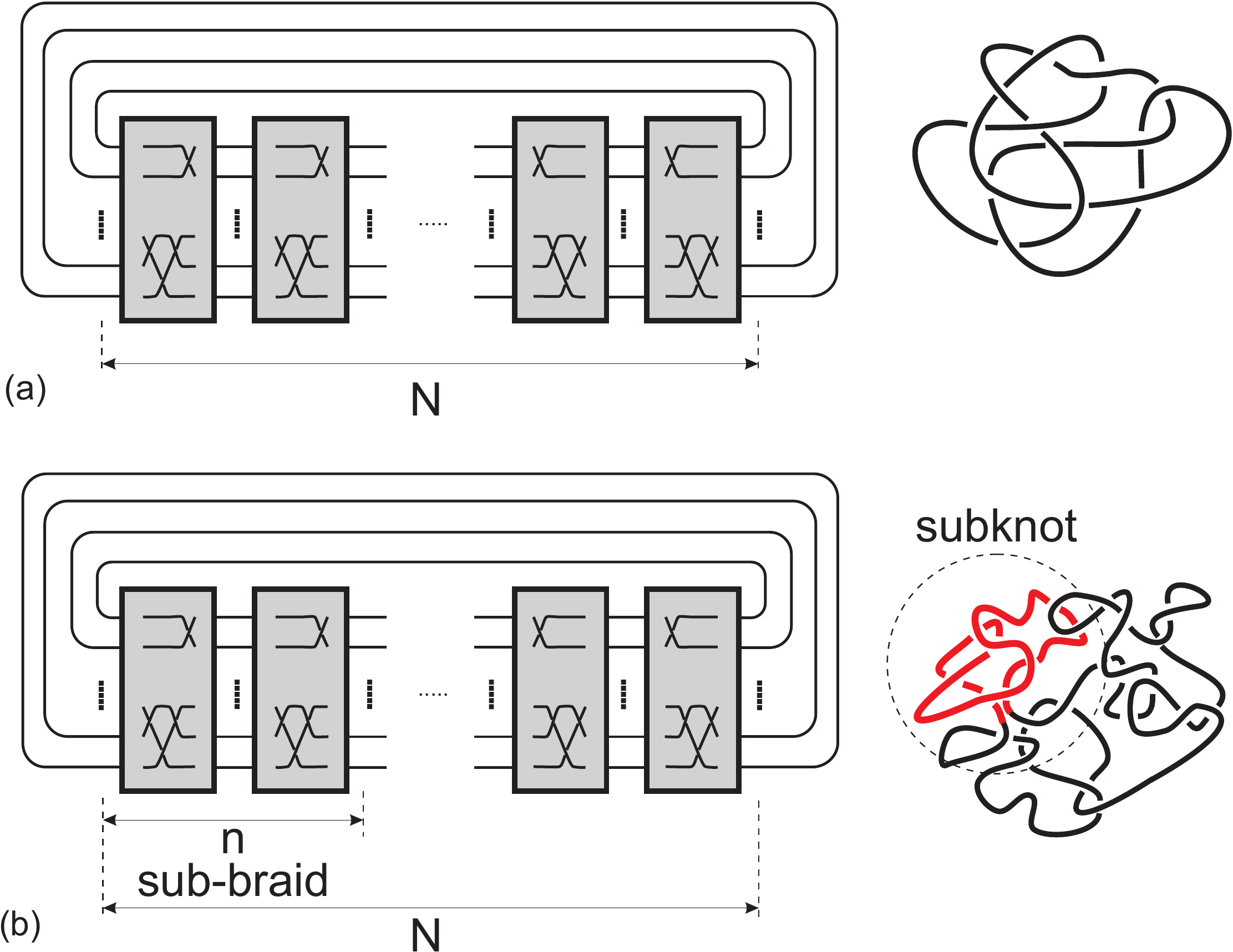}
\caption{Schematic representation of knots by braids: a) unconditional random distribution of black
boxes produces a very complex knot; b) conditional distribution implies the whole knot to be
trivial, which imposed strong constraints on complexity of any subpart of the braid.}
\label{fig:braids}
\end{figure}

Our topological problem to determine the complexity of a subloop in a globally trivial collapsed
polymer chain allows natural interpretation in terms of BB. Suppose the following imaginative
experiment. Consider the phase space $\Omega$ of all topological states of densely packed knots on
the lattice. Select from $\Omega$ the subset $\omega$ of trivial knots. To simplify the setting,
consider a knot represented by a braid, as shown in the \fig{fig:braids}, where the braid is
depicted by a sequence of uncorrelated "black boxes" (each black box contains some number of over--
and under--crossings). If crossings in all black boxes are identically and uniformly distributed,
then the boxes are statistically similar. Cut a part of each braid in the subset $\omega$, close
open tails and investigate the topological properties of resulting knots. Just such situation has
been qualitatively studied in \cite{gns}, where the crumpled globule concept was formulated mainly
on the basis of heuristic scaling arguments. The CG hypothesis states the following: if the whole
densely packed lattice knot is trivial, then the topological state of each of its "daughter" knot
is almost trivial.

It has been shown \cite{nech_vas} that the computation of the knot complexity in the braid
representation depicted in the \fig{fig:braids} can be interpreted as the computation of the
highest eigenvalue of the product of noncommutative matrices designated by black boxes.

To proceed, consider first the typical (unconditional) complexity of a knot represented by a
sequence of $N$ independent black boxes. This question is similar to the growth of the logarithm of
the largest eigenvalue, $\lambda$, of the product of $N$ independent identically distributed
noncommutative random matrices. According to the Furstenberg theorem
\cite{furstenberg1963noncommuting}, in the limit $N\gg 1$ one has
\be
\ln \lambda(N) \sim \gamma N,
\label{eq:uncond}
\ee
where $\gamma={\rm const}$ is the so-called Lyapunov exponent. Being rephrased for knots, this
result means that the average knot complexity, $\varkappa$, understood as a minimal number of
crossings, $M$, necessary to represent a given knot by the compact knot diagram, extensively grows
with $M$, i.e. $\varkappa \sim M$. In the ordinary globule, for subchains of length $N^{2/3} < s <
N$, the typical number of crossing, $M$, on the knot diagram grows as $M\sim s^2$, leading to the
scaling behavior
\be
\varkappa \sim s^2
\label{eq:vark1}
\ee
for the knot complexity $\varkappa$. This is perfectly consistent with the well known fact: the
probability of spontaneous unknotting of a polymer with open ends in a globular phase is
exponentially small. Following the standard scheme \cite{knots-kardar,virnau}, we characterize the
knot complexity, $\varkappa$, by the logarithm of the Alexander polynomial, $\ln [{\rm Al}(t=-1.1)
{\rm Al}(t=-1/1.1)]$, i.e. we set $\varkappa=\ln [{\rm Al}(t=-1.1)  {\rm Al}(t=-1/1.1)]$. As it
seen from \fig{fig:02}, the conjectured dependence $\ln {\rm Al}(t=-1.1)\sim
s^2$ is perfectly satisfied for ordinary (knotted) globule.

Consider now the \emph{conditional} distribution on the products of identically distributed black
boxes. We demand the product of matrices represented by black boxes to be a unit matrix. The
question of interest concerns the typical behavior of $\ln \tilde{\lambda}(n,N)$, where
$\tilde{\lambda}(n,N)$ is a sub-chain of first $n$ matrices in the chain of $N$ ones. The answer to
this question is known \cite{nechaev1991limiting}:
if $n=cN$ ($0<c<1$ and $N\gg 1$), then
\be
\ln \tilde{\lambda}(n=cN,N) \sim \sqrt{n} = \tilde{\gamma}(c) \sqrt{N}
\label{eq:cond}
\ee
where $\tilde{\gamma}(c)$ absorbs all constants independent on $N$. Translated to the knot
language, the condition for a product of $N$ matrices to be completely reducible, means that the
"parent" knot is trivial. Under this condition we are interested in the typical complexity
$\tilde{\varkappa}$ of any "daughter" sub-knot represented by first $n=cN$ black boxes.

Applying the \eq{eq:cond} to the knot diagram of the unknotted globule, we conclude that the typical
conditional complexity, $\tilde{\varkappa}$ expressed in the minimal number of crossings of any
finite sub-chain of a trivial parent knot, grows as
\be
\tilde{\varkappa}\sim \sqrt{s^2} \sim s
\label{eq:vark2}
\ee
with the subchain size, $s$. Comparing \eq{eq:vark2} and \eq{eq:vark1}, we conclude that
subchains of length $s$ in the trivial knot are much less entangled/knotted than subchains
of same lengths in the ``unconditional'' structure, i.e. when the constraint for a parent knot to
be trivial is relaxed. Indeed, this result is perfectly supported by \fig{fig:02} which show linear grows of $\tilde{\varkappa} = \ln [{\rm
Al}(t=-1.1)\,{\rm Al}(t=-1/1.1)]$ with $s$ for the unknotted globule, while quadratic grows for
the knotted globule.


\clearpage
\section*{Supplemental Information}
\setcounter{figure}{0} \renewcommand{\thefigure}{S\arabic{figure}}

\subsection*{Scaling arguments for $P_c(s)$ in fractal and equilibrium globules, and random walks}

Recall that the contact probability $P_c(s)$ is the probability that two monomers separated by the
distance $s$ along the chain come close enough in the space to make a contact. The probability of
forming a contact in a globular state with a constant density can be roughly estimated as an
inverse of a volume $V(s)$ in which two ends of a subchain of legnth $s$ reside: $P_c(s)\sim 1/V(s)
\sim 1/R_G^D(s)$. Here, $R_G(s)$ is the gyration radius of the chain fragment of $s$ monomers. For
a curve with dense space-filling subchains, the volume of a subchain is proportional to the number
of monomers $V(s) \sim s$, yielding $R_G(s) \sim s^{1/3}$ and $P_c(s) \sim s^{-1}$. This decay has
been discussed in the literature \cite{lieberman}, so the above naive derivation of the
$s^{-1}$--law pursues mainly illustrative aims.

For a random walk, subchains are gaussian, and therefore have average size $R_G(s) \sim s^{1/2}$.
Therefore, two ends of a subchain are located within the volume $V(s) \sim R_G^3 \sim s^{3/2}$.
Probability that the two ends of the subchain of length $s$ are in contact is approximately
inversely proportional to the volume in which the subchain resides $P_c(s) \sim \frac1{V(s)} \sim
s^{-3/2}$. In the equilibrium globule, subchains of sizes $s < N^{2/3}$  behave as random walks,
and thus have the contact probability $P_c(s) \sim \frac1{V(s)} \sim s^{-3/2}$. For subchains longer
than $s =N^{2/3}$, subchains start reflecting from the boundaries of the confinement, positions of
the two ends become uncorrelated in the volume of the entire chain, and thus $P_c(s)$ exhibits a plateau.

\begin{figure}[ht]
\includegraphics[width=8cm]{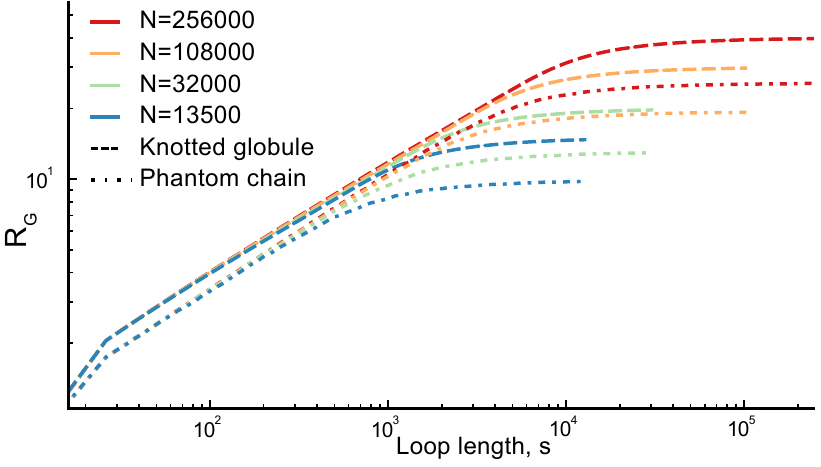}
\caption{Gyration radius for subchains of knotted globules and confined phantom chains; similar to Fig. \ref{fig:03}a.}
\label{fig:Phantom}
\end{figure}

We note that $P_c(s)$ and $R_G(s)$ for the equilibrium globule are qualitatively similar with $P_c(s)$ and $R_G(s)$ for a phantom chain confined to the same box; plots for $Rg(s)$ are shown in the figure \ref{fig:Phantom}.
However, for a phantom chain confined to the same volume, the plateau in $P_c(s)$ and $R_G(s)$ starts earlier, because the
chain is repelled by the walls and thus has a higher density in the center and a lower density next to the walls, which corresponds to an effectively smaller confining volume.
Repulsion of a phantom chain by confining walls is well-studied, and analytic expression for spatial density can be obtained; for details see \cite{GKh-book}.

\subsection*{Relation between $P_c(s)$ and $R_G(s)$}

\begin{figure}[ht]
\includegraphics[width=8.5cm]{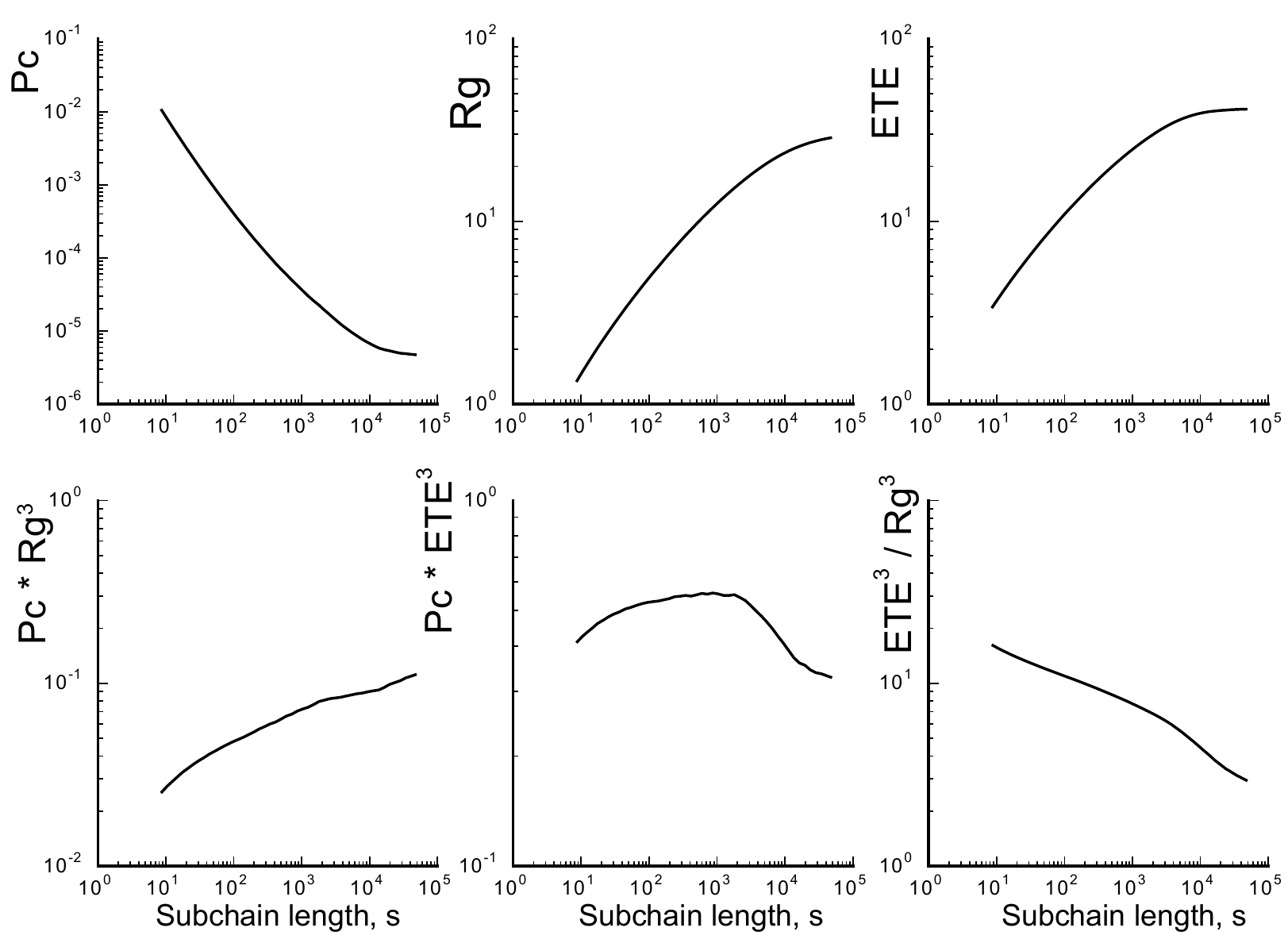}
\caption{(top) Average contact probability, gyration radius of subchains, and end-to-end distance for $256\,000$ long unknotted globules. (bottom) $P_c(s)$, renormalized by gyration radius (left), end-to-end distance(middle). Cube of the ratio of $R_g$ to the end-to-end distance. (right)}
\label{fig:renorm}
\end{figure}

Contact probability, $P_c(s)$, it not a dimensionless quantity, and contains information both about the size of a subchain, ($R_G(s)$), and the internal structure of the subchain. To distinguish between the two effects on $P_c(s)$, we multiplied the $P_c(s)$ plot by the approximate subchain volume, $R_G(s)^3$, thus accounting for the effect of subchain size. However, we found that the results were drastically different dependent on whether we used gyration radius $R_g(s)$, or average distance between monomers separated by $s$ ("End-to-end distance", ETE), as a measure of subchain size. Specifically, $Pc(s)$ renormalized using gyration radius is monotonically increasing with chain size, while $Pc(s)$ renormalized using end-to-end distance has a maximum for intermediate-size subchains (see Fig. \ref{fig:renorm}).  This discrepancy is likely explained by the lack of well-defined scaling relations in our system. These results highlight that extra care should be taken when interpreting slopes of $P_c(s)$ and $R_G(s)$ curves as indicators of scaling relations. However, we emphasize that the contact probability is relevant for studies of chromosomes, as it is related to the frequency of interactions between genomic elements, and thus can be measured directly using Hi-C.

\subsection*{Effects of confinement}

\begin{figure}[ht]
\includegraphics[width=8cm]{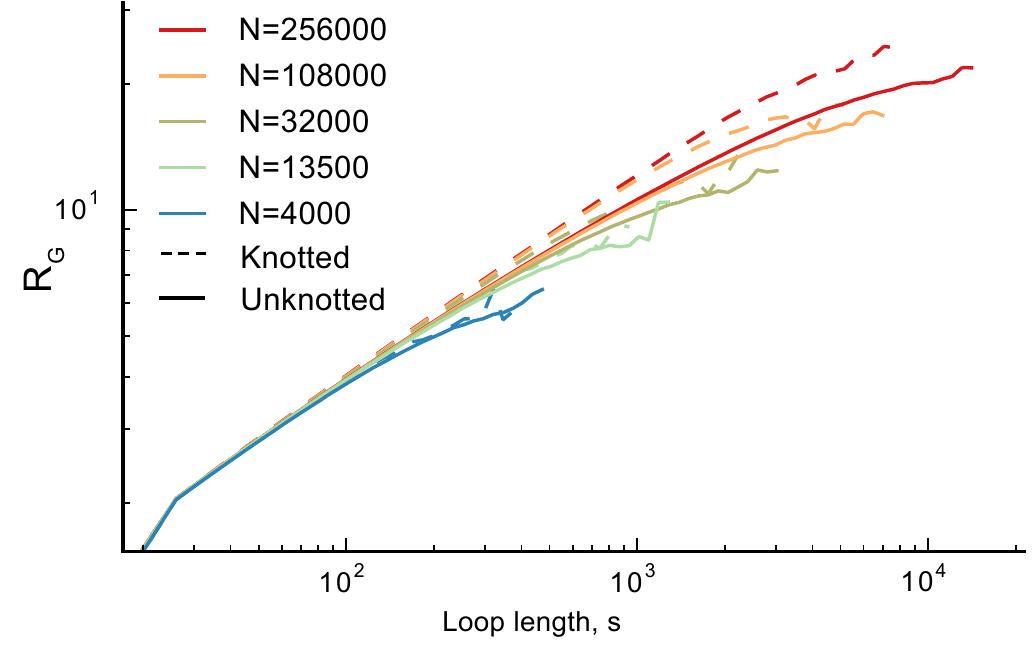}
\caption{The same plot as in Fig. \ref{fig:03}a, but evaluated for loops which are not in contact with the boundary. A loop was defined to be in contact with the boundary if at least one monomer of the loop occupied the site directly adjacent to the box.}
\label{fig:not_touching}
\end{figure}

Most of the larger loops in our system are in contact with the confining boundary. However, if we only focus on the loops which are not touching the boundary, we can still see the difference between the knotted and the unknotted globules.  \fig{fig:not_touching} shows the $R_G(s)$ plots for different chains, only focusing on the loops which are not in contact with the boundary. Note that $R_G(s)$ plots terminate at $s \le N$ because most of the large loops have at least one of the monomers in contact with the boundary.

We note that gyration radius, $R_G$ is characterizing the second moment of the spatial distribution of monomers in a subchain: $R_G = <(x - \bar x)^2> + <(y - \bar y)^2> + <(z - \bar z)^2> = <(r-c.o.m.)^2>$, where c.o.m. denotes center of mass. Here we study if higher moments of the distribution to the distances to the center of mass are different in subchains of the knotted and unknotted globules. However, we can only compare subchains of the same average size $R_G$, as the effect of confinement on the distribution of distances to the center of mass is dependent on the $R_G$. For subchains in a confined volume, spatial distances exceeding the confining box size are not possible; this cuts off tails of the distribution of distances between monomers and makes the distribution non-gaussian, especially when $R_G$ reaches the size of confining box.

\begin{figure}[ht]
\includegraphics[width=8cm]{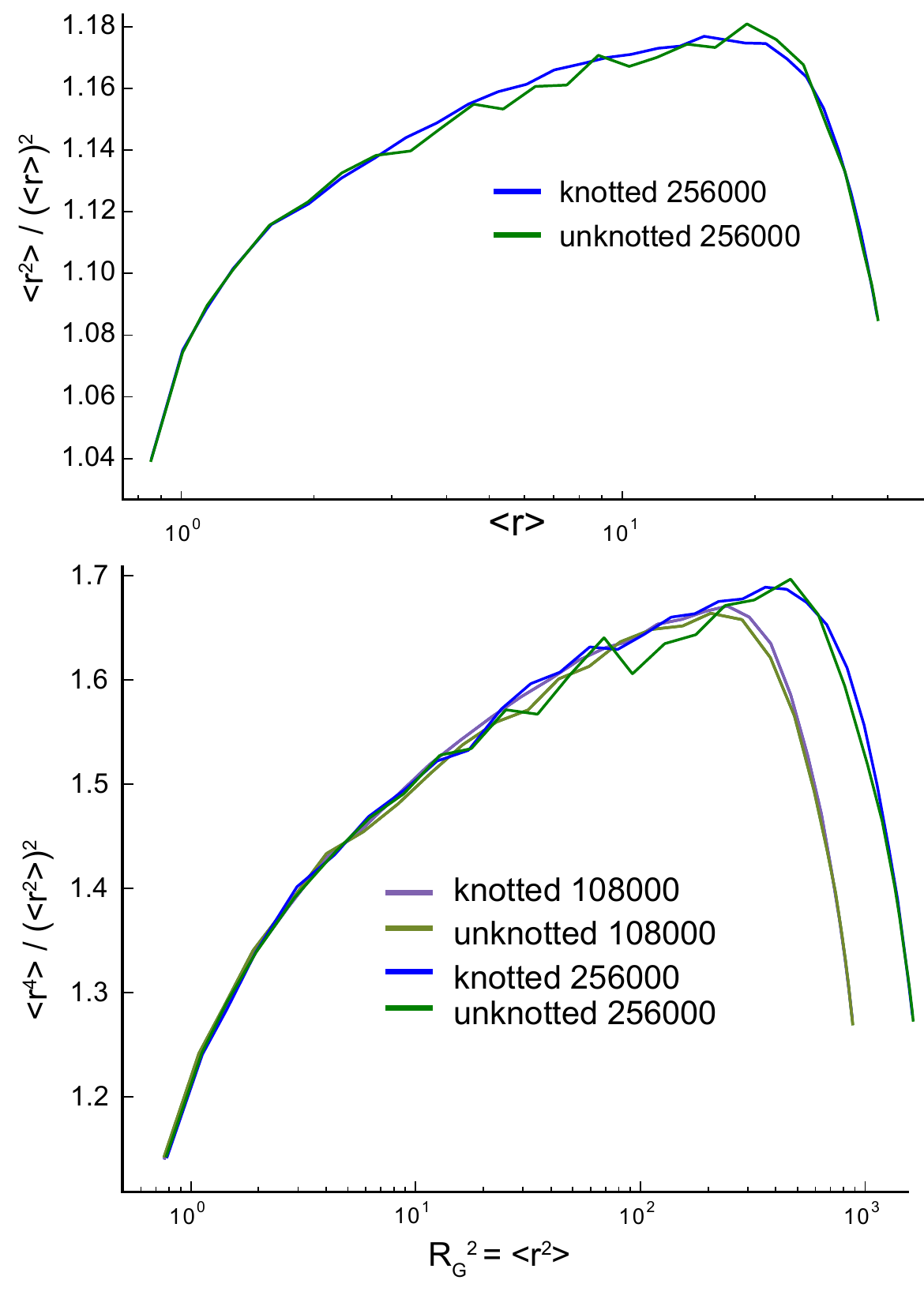}
\caption{(top)Dependence of the normalized second moment of spatial distribution, $<r^2> / (<r>)^2$ on the first moment,$<r>$; $r$ is distance to the center of mass of a subchain.  (bottom)  Dependence of the normalized fourth moment of spatial distribution, $<r^4> / (<r^2>)^2$ on the second moment,$<r^2> = R_G^2$. }
\label{fig:kurtosis}
\end{figure}

To test whether knotted and unknotted globules have different distribution of distances to the center of mass, we plotted normalized moment ratios this distribution. Gyration radius is a second moment, $<(r - \bar r)^2>$. We plotted normalized second moment, $<r^2> / <r>^2$ as a function of $<r>$, and normalized fourth moment as a function of the second. The latter is analogous to the kurtosis of the distribution. Since subchains of the same $R_G$ or $<r>$ are expected to experience the same boundary effects, then any differences in the plots would be indicative of a non-trivial structure of subchains in unknotted globules, as compared to knotted. Surprisingly, we find no difference in how the normalized higher moments of the distribution depend on the mean. This is indicative of the fact that subchains of the same spatial size in knotted and unknotted globules have comparable  spatial structure ( Fig. \ref{fig:kurtosis}).

\subsection*{Topological properties of subchains}

To obtain knotting diagrams, we evaluated $\varkappa$ for all loops in a given globule. We then divided all loop start positions and loop length in the equal number of bins ($500$ monomer bins for $32\,000$ globules, $1\,000$ for $108\,000$). For each bin of loop lengths and loop start, we evaluated average $\varkappa$ for all loops contributing to a given point, and record the average value of the knotting complexity. We then display average loop complexities on a map, using "jet" colormap, and showing $\sqrt{\varkappa}$ to make unknots easily distinguishable. White regions of the map correspond to the pairs of regions of the globule which formed no loops at all.

\begin{figure}[ht]
\includegraphics[width=8cm]{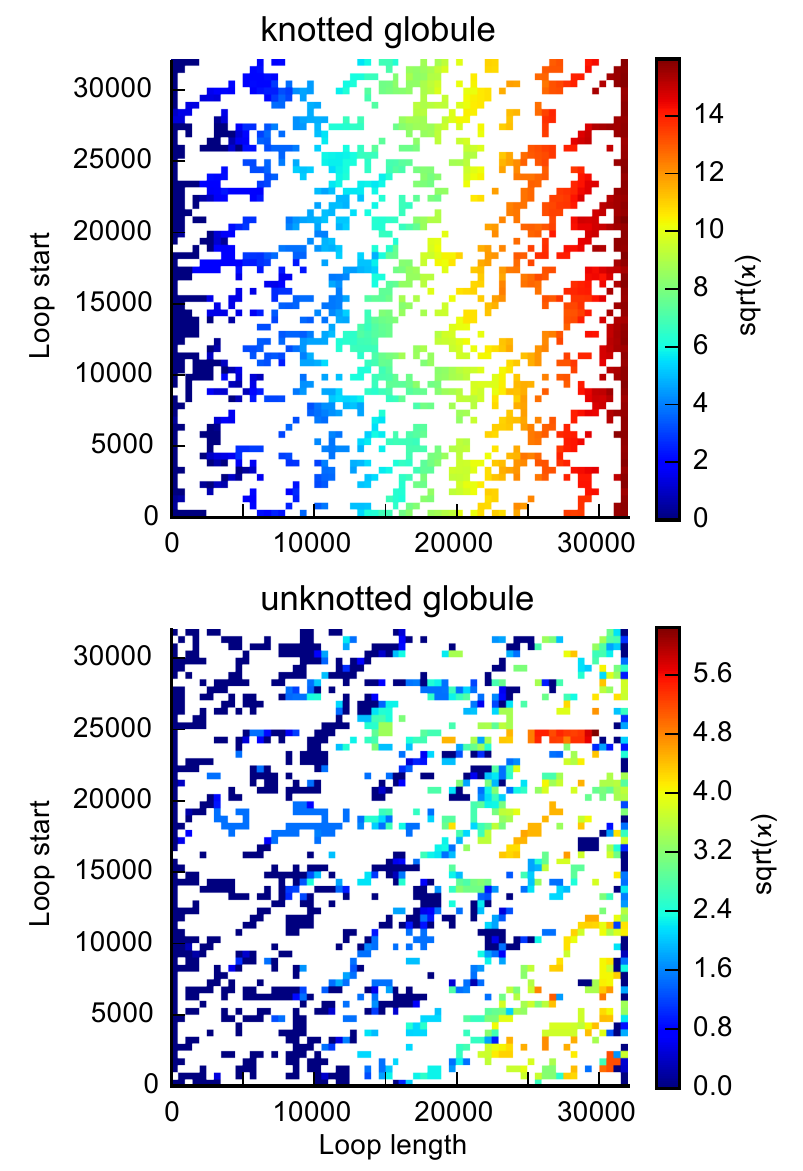}
\caption{Matrices of loop knot complexity for a single $32\,000$-long knotted globule (top), and unknotted globule (bottom). Color shows $\sqrt{\varkappa}$ to better highlight unknots. Note that the color scale is different for the two images. }
\label{fig:slipknots32}
\end{figure}

Figure \ref{fig:slipknots32} shows matrices of knot complexity for 32000-long knotted and unknotted globules. As we can see, knotting of loops in the unknotted globule is highly variable, and for any loop length it ranges from completely unknotted to highly knotted. On the contrary, for loops in the knotted globule, knot complexity depends strongly on loop length, and is not very variable for a given loop length (color forms uniform vertical stripes).

We note that in both types of globules, loops longer than $N/2$ are more knotted than shorter loops. Each contact in the unknotted globule separates the polymer into two loops: one loop that is larger than $N/2$ and the other that is smaller. If the two loops are unlinked, they would both have to be unknotted. However, if they are linked, then their knotting complexities are not necessarily equal; in fact, it often happens that the shorter loop would be unknotted and the longer loop would be knotted. For example, in a $108\,000$-long globule, $98.8 \%$ loops of length $500$ to $1000$ are unknotted, while $58\%$ of the matching loops of length $107\,000$ to $107\,500$ are knotted with the average knot complexity $\varkappa = 14$.

\begin{figure}[ht]
\includegraphics[width=8.3cm]{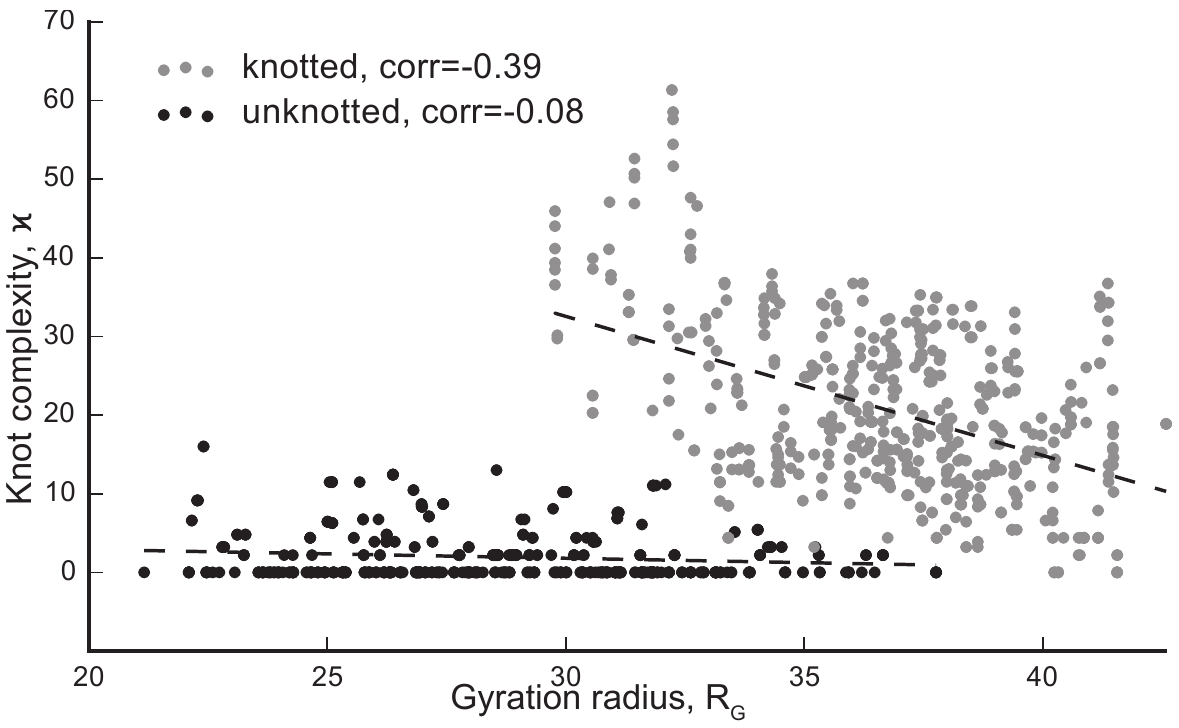}
\caption{A scatter plot of $R_G(s)$ and knot complexity for $s=20\,000-22\,000$ loops in $N=256\,000$ knotted and unknotted globules.
}
\label{fig:3c_suppl}
\end{figure}

\section*{Polymer Simulations and Analyses}

\subsection*{Initial conformation}

We note that since simulations are preformed at equilibrium, initial conformation is irrelevant to the conclusions made in the paper, and is presented for consistency and responsibility only. We initialize polymer simulations from two different starting conformations: growing polymer ring, and polymer packed in a grid. A growing polymer ring on a cubic lattice was created from a small 4-monomer ring by incrementally adding two monomers to it. At a first step, a 4-monomer ring was placed in the center of the box used for simulations. Then, one bond was chosen randomly. We then tried to extend the polymer at the chosen bond by two monomers, by making the bond into a kink. To do this, we considered another bond, obtained by shifting this bond by distance $1$ in a random direction perpendicular to the bond (chosen out of 4 possible directions). If both locations of the shifted bond were free, the polymer was extended to incorporate this bond. For example, if the bond was going in $+z$ direction: $... \to (0,0,0) \to (0,0,1) \to ...$, and we attempted to grow it in the $-y$ direction (chosen randomly out of $+x, -x, +y, -y$), we would check positions $(0,-1,0)$ and $(0,-1,1)$. If both of them were free, the polymer sequence would be changed to $... \to (0,0,0) \to (0,-1,0) \to (0,-1,1) \to (0,0,1) \to ...$. If at least one of them were occupied, selection of a random bond would be repeated. The process was repeated till the polymer grew to the desired length.

Polymer densely packed in a grid was created as described below. Here $\{\}$ indicates the largest unit, $[~]$ indicates a smaller unit, and $(~)$ indicates a monomer
\begin{multline*}
\{[(1,1,1) \to (2,1,1) \to ... \to (N-2,1,1) \to \\ \hspace{-2.5cm} (N-2,2,1) \to ... \to (1,2,1)]
\to \\ \hspace{-3cm} [(1,3,1) \to ... (1,4,1)] \to ... \to \\ \hspace{-1.8cm} [(1,N-3,1) \to ...
\to (1,N-2,1)] \to \\ \hspace{-0.95cm} [(1,N-2,2) \to ... \to (1, N-3,2)] \to ... \to \\
\hspace{-3.1cm} [(1,2,2) \to ... \to (1,1,2)] \} \to \\ \hspace{-1cm}  \{[(1,1,3) \to ... \to
(1,1,4)]\} \to ... \to ((a,b,c)) \\
 \end{multline*}
after which it was crawling back to $(1,1,1)$ along the $(x,y)$ plane, and then through the $(0,y,z)$ plane) when the distance to the end was barely enough to make the return path.

All simulations in the paper were initialized from a growing polymer ring; a control initialized from a polymer packed in a grid was performed for $N=108\,000$, and showed $R_G(s)$ and $P_c(s)$ within 1\% of that for a system initialized from the growing ring.

\subsection*{Polymer systems}

We performed polymer simulations as described in \cite{gr-krem1}, considering 6 system
sizes: $256\,000$ monomers in a $80\times 80\times 80$ box, $108\,000$ monomers in a $60\times 60\times
60$ box, $32\,000$ monomers in a $40\times 40 \times 40$ box, $13\,500$ monomers in a $30\times 30\times
30$ box, 4000 in a $20\times 20\times 20$ box and 2048 in a $16\times 16 \times 16$ box. Simulation
times, measured in $\log{t} / \log{N}$ are shown in \fig{fig:07b}. We performed $1000$ replicas for
unknotted globules up to $13\,500$ monomers, $400$ replicas for $32\,000$ monomers, $30$ replicas for
$108\,000$ monomers and $20$ replicas for the largest system.

\begin{figure}[ht]
\includegraphics[width=8cm]{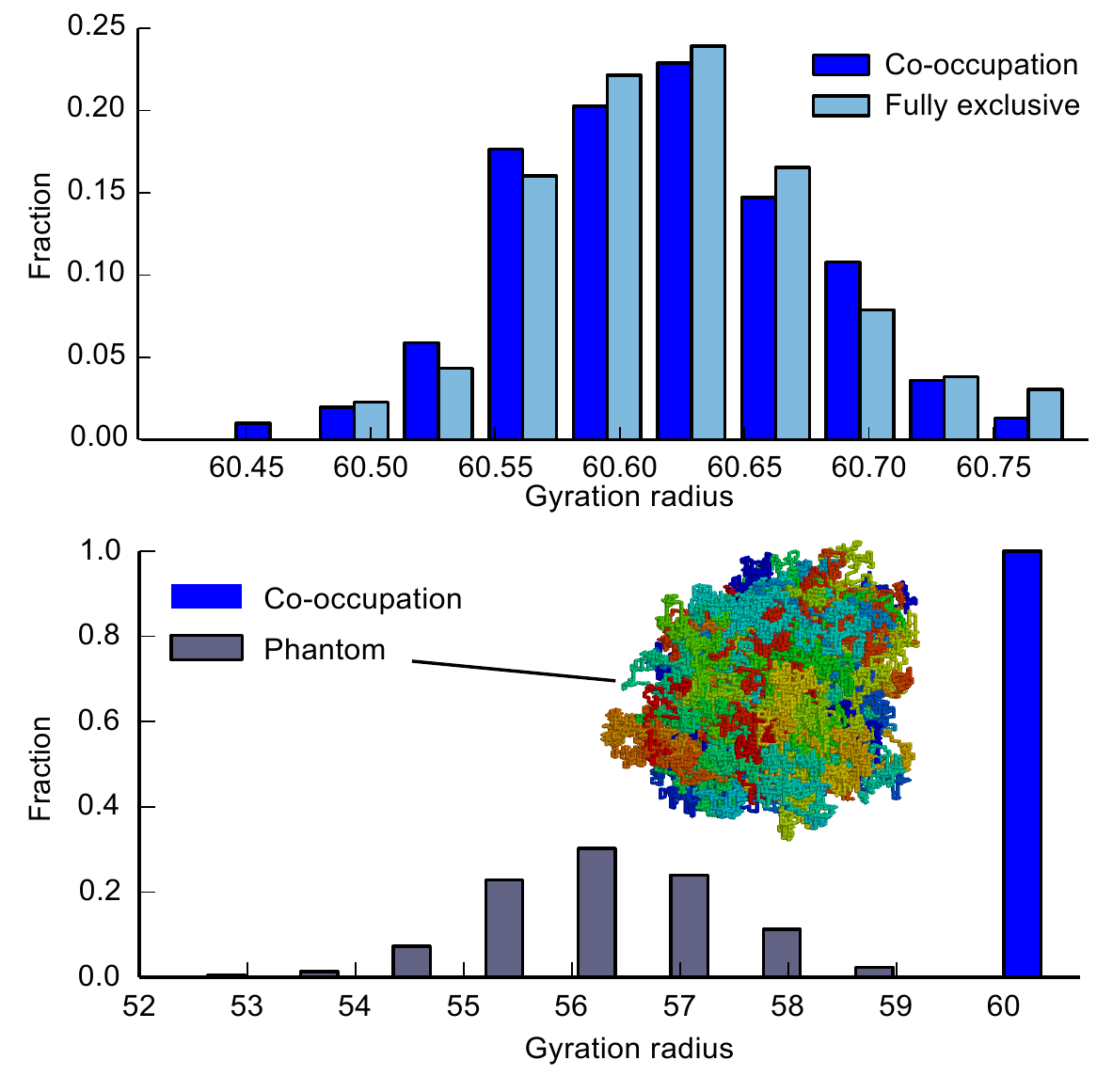}
\caption{(top) $R_G$ of the whole knotted globule with partial co-occupation of
lattice cites (i.e. how the knotted globule was simulated; shown in blue) is compared to the $R_G$ of the knotted globule for which full strength excluded volume interactions were introduced (light blue).
(bottom) $R_G$ of the knotted globule with co-occupation (blue) is compared to the $R_G$ of the phantom chain (gray). Note that the X scale is different compared to the above plot. Insert shows a sample conformation of the phantom chain. Note that it does not have a sharp boundary, unlike knotted or unknotted globules (\fig{fig:01})}
\label{fig:Co-occupation}
\end{figure}

We simulated topologically relaxed system by allowing co-occupation of the same lattice site by two monomers with a probability of $0.002$. To test whether this procedure introduces any change in spatial properties of the polymer, we turned on full strength excluded volume interaction in $N=108\,000$ equilibrated topologically relaxed globules, and simulated them for additional $1\cdot 10^7$ MC steps per monomer. This process resolved all co-occupation events after $1\cdot 10^3$ MC steps per monomer, and simulated the globule for $10^4$ times longer. We then compared gyration radii $R_G$ of the whole chain with co-occupation and without co-occupation, and found no noticeable difference in the full-chain $R_G$ (\fig{fig:Co-occupation} top). Note that if the chain was made completely phantom (i.e. 100\% co-occupation), resulting shift of $R_G$ would be much more dramatic, and $R_G$ would decrease much more (\fig{fig:Co-occupation}, \ref{fig:Phantom}), producing visually different chains.

\begin{figure}[ht]
\includegraphics[width=8cm]{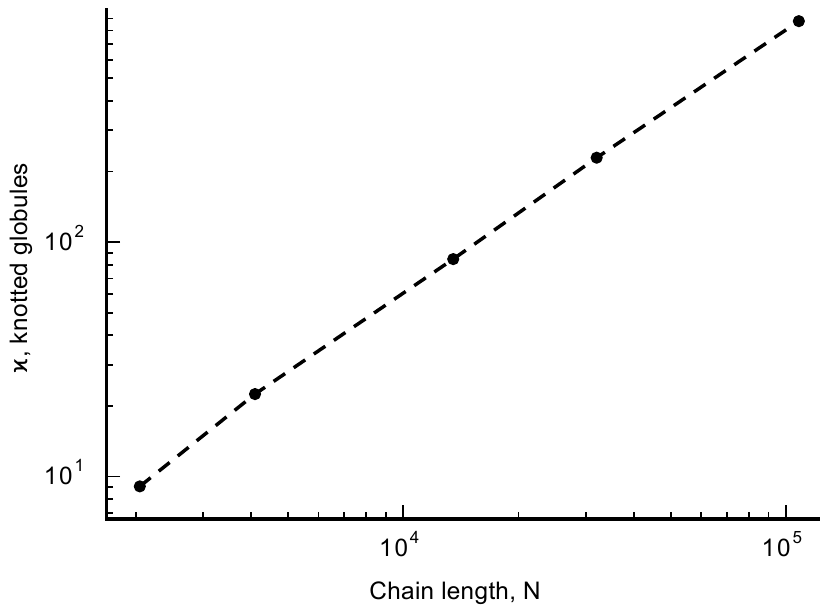}
\caption{Average $\varkappa$ for knotted globules
of sizes $2048$, $4000$, $13\,500$, $32\,000$, and $108\,000$.}
\label{fig:AlexKnotted}
\end{figure}

The knot complexity $\varkappa$ of the whole chain for topologically relaxed polymer system is increasing sharply with chain size. For $N=256\,000$, calculation of Alexanders polynomial for the whole chain was computationally unfeasible; For $N=2048-108\,000$, average $\varkappa$ in a knotted globule is shown in \fig{fig:AlexKnotted}.

\subsection*{Equilibration}

\begin{figure}[ht]
\includegraphics[width=8.5cm]{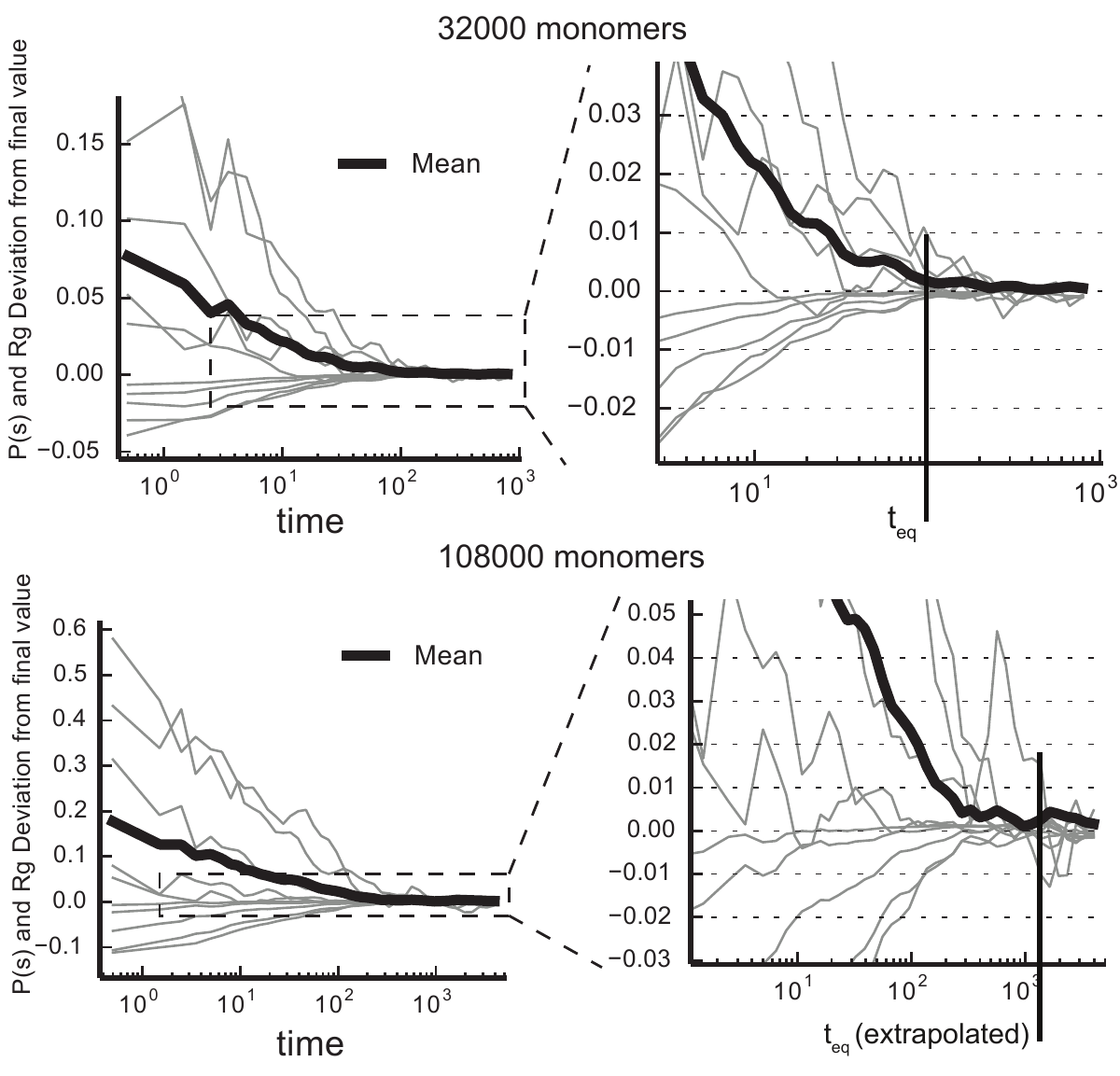}
\caption{Contact probability $P_c$ and gyration radius $R_G$ shown as a function of time for five distinct values of subchain length $s$ for a $32\,000$-long globules (top) and $108\,000$-monomer globules (bottom). Thick black line labelled as "mean" denotes average absolute deviation of 10 curves from zero. Plots were divided by the mean value of the last 50\% of the trajectory, and then 1 was subtracted to show relative deviation from the final value. Equilibration time for $N=32\,000$ monomers was estimated as described in the text and equals $N^{1.55}$. Extrapolated time for $N=108\,000$ monomers is calculated from a more permissive $N^{1.6}$ dependence.}
\label{fig:07}
\end{figure}

\begin{figure}[ht]
\includegraphics[width=8.4cm]{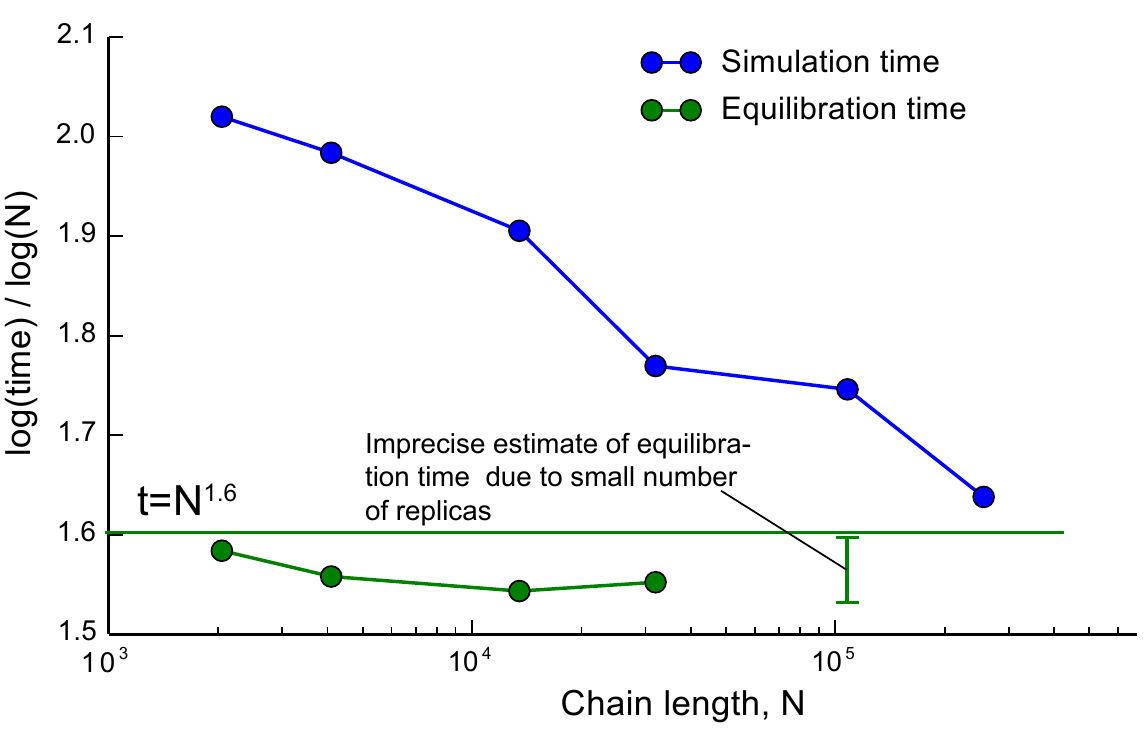}
\caption{Time of simulations and time of equilibration shown for unknotted globules of various sizes. Time is measured in logarithm base N, where N is number of monomers. Equilibration time for $N=108\,000$ cannot be precisely estimated because fluctuations in the average $P_c(s)$ and $R_G(s)$ exceeds 1\% due to small sample size . Upper and lower values of the range shown for $N=108\,000$ denote the time when all 10, or 9 out of 10 observables fall below 1\% relative deviation. Note that in \fig{fig:07} time points are logarithmically spaced, and hence time points for longer times aggregate information over longer time spawns. Therefore fluctuations due to the small number of replicas decrease with time, and allow us to put an upper boundary on the equilibration time for $N=108\,000$.}
\label{fig:07b}
\end{figure}

To quantify equilibration of the system, we evaluate two observables: $P_c(s)$ and $R_G(s)$ for subchains of length $s = N ^ \alpha$, where $\alpha$ was equal  $0.2, 0.35, 0.5, 0.65, 0.8$. We performed simulations of small systems ($N \le 32\,000$) for long time (\fig{fig:07b}) with $400-1000$ replicas for each system size. We then quantitatively defined equilibration as a point where all 10 $R_G(t)$ and $P_c(t)$ curves defined above fall within 1\% deviation from the final $P_c(s)$ or $R_G(s)$ (\fig{fig:07}). The final value for the plot was defined as an average over the last half of the trajectory. We note that equilibration time have always occurred within the first 10\% of the trajectory (20\% for $N=108\,000$), allowing comparison to the last half.

We then plot $log(t_{eq})$ as a function of $s$, and obtain that the system equilibrates for $t < N^{1.6}$ steps per monomer for all systems up to $32\,000$ monomers. Specific values of the power $\alpha = \log(t)/\log(N)$ were: $1.583$, $1.558$, $1.543$, $1.552$ for $2048$, $4096$, $13\,500$, $32\,000$ monomers respectively (\fig{fig:07b}). We then extrapolate the  $N^{1.6}$ time to longer chain lengths, $N =128\,000,~256\,000$, as we cannot obtain enough statistics to estimate $P_c(s)$ to less than 1\% at that chain lengths. A longer, $128\,000$-monomer system was simulated for $6\, N^{1.6}$, while $256\,000$ system was simulated for $2.2\, N^{1.6}$. Based on the $P_c(s)$ and $R_G(s)$ plots, one can see that the $108\,000$-monomer system is equilibrated according to our criteria; however, we cannot estimate equilibration time precisely because of  fluctuation in in $R_G(s)$ and $P_c(s)$ plots for different time points (\fig{fig:07}) due to a small number of replicas. Consequently, we cannot use this system size for extrapolation. Thus we conclude that $108\,000$-monomer system is equilibrated, while $256\,000$-monomer system is equilibrated based on the extrapolation.

\subsection*{Contact probability and gyration radius analyses}

We define two monomers to be in contact if they occupy two adjacent sites on a cubic lattice. To plot $P_c(s)$, we first divide all inter-monomer separations in bins of logarithmically increasing sizes, starting at $6$ and going with a step of $1.1$, rounding to the nearest even integer and removing duplicates. This yielded a sequence of $6, 8, 10,  ... , x, 1.1  x, 1.1^2  x, ... ,N$.

To define $R_G(s)$ for loops, we divided all distances to similar logarithmically space bins. We then considered all contacts which occur at that distance, and sampled a fixed number of loops from that set. If no contacts were present at a given bin of separations, we ignored this conformation; however, this happened in a fraction of a percent of cases. Averages were performed over replicas and over time. To obtain plots in the main text, these quantities were averaged over the last half of the trajectory for each run; for time-dependence plots they were averaged over logarithmically-spaced time bins with a coefficient of $1.2$.

\subsection*{Calculation of Alexander polynomial}

Finding Alexander polynomial for large rings could be a computationally intensive task. Here, we used the following strategy to do it.

First, we attempted to simplify the polymer ring using the previously used algorithm. We chose a pair of neighboring polymer bonds and attempted to remove a particle between them, replacing two bonds with one. We then checked if any of the other bonds are crossing the triangle formed by the two original bonds and the new bond. If none of the bonds were crossing this triangle, we accepted the replacement. We then continued choosing bonds until no further simplification could be made. This part was thoroughly tested, and we detected no differences in Alexander polynomial of the original and simplified polymer.

Second, we used the code adjusted from  \cite{kolesov2007protein} to calculate the value of Alexander polynomial at -1.1. The current code is at http://bitbucket.org/mirnylab/openmm-polymer in a knots folder.

\begin{figure*}[ht]
\includegraphics[width=18cm]{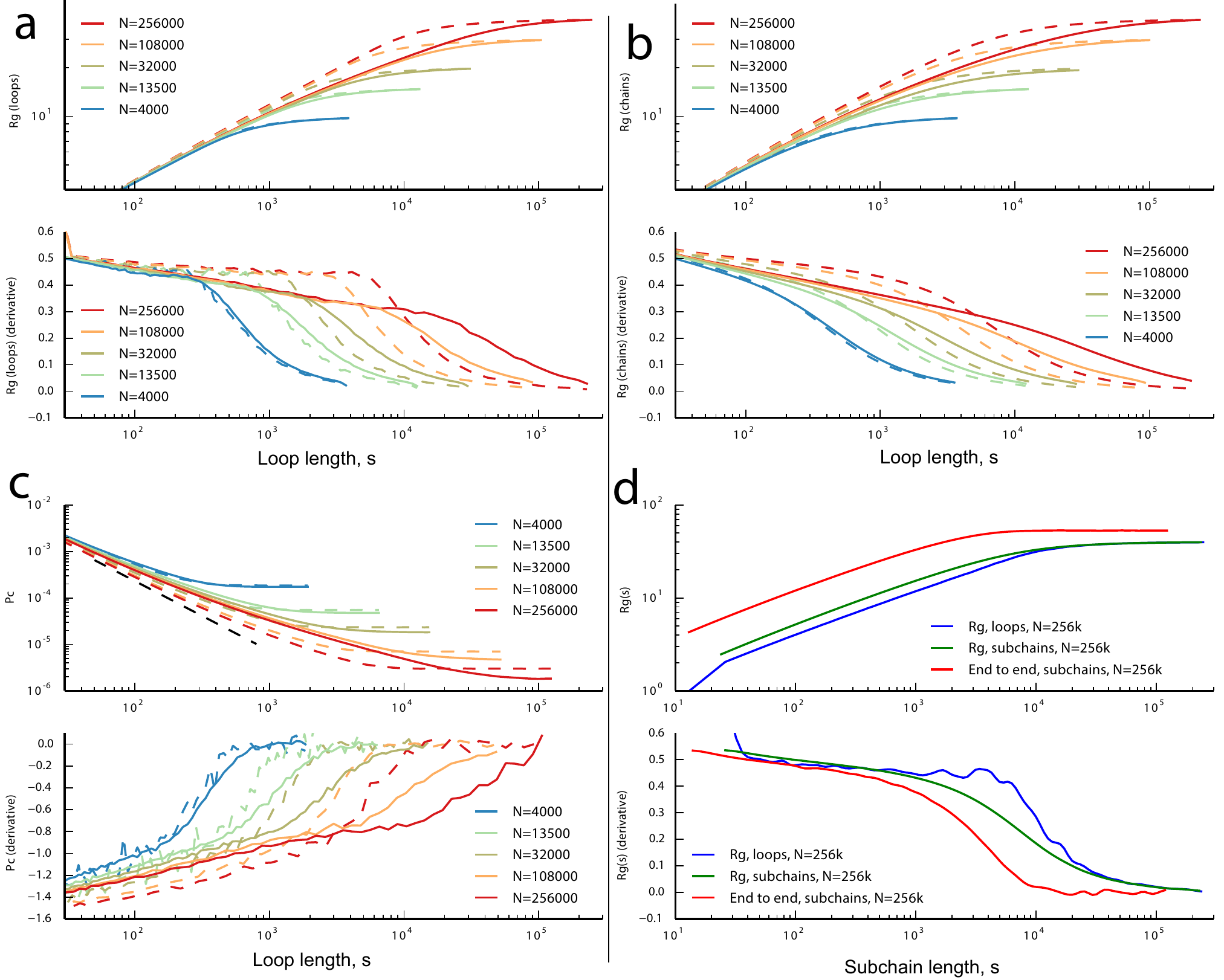}

\caption{In this figure we explore whether scaling regimes for $R_G(s)$ and $P_c(s)$ can be established. We compute a slope in the log-log coordinates as $d \log{R_G(s)} / d \log(s)$ and plot it as a function of $\log(s)$.
As in the main text, solid line corresponds to the knotted globule, and dashed line - to the unknotted. Each of the four panels has two plots: the top plot is the quantity of interest ($R(s)$ or $P_c(s)$), and the bottom plot is its slope. {\bf (a)} $R_G(s)$ for loops, similar to Fig. 3. {\bf (b)}  $R_G(s)$ evaluated for subchains. Note that both for the knotted and unknotted globule, all transitions between scaling regimes are more smooth.  {\bf (c)} The contact probability $P_c(s)$.  The effect of confinement is more evident here affecting the scaling of $P_c$ even for $s=100$ for the knotted globule of $N=256\,000$. Moreover, $P_c(s)$ has a plateau at $s \sim N$ for the knotted globule, which is not observed for corresponding $R_G(s)$. {\bf (d)} Three quantities characterizing the spatial size of subchains in the knotted globule: $R_G$ for loops, $R_G$ for subchains, and  the end-to-end distance $R$.  For the same subchain length $s$, $R_G(s)$ for loops is smaller, and thus it starts to feel the effects of confinement at larger $s$, showing a clear $R_G(s) \sim s^{0.5}$ scaling regime up to several thousand monomers. $R_G(s)$ evaluated for subchains is larger, and the effect of constraints starts to manifest itself  for smaller values of $s$. Finally, the end-to-end distance is the largest of the three,  which makes it feel the confinement at even smaller $s$. The end-to-end distance shows a distinct plateau at $s \gtrsim 10\,000$. A similar plateau is seen in the $P_c(s)$ plots.}
\label{fig:fullpage}
\end{figure*}

\end{document}